%% file: article.tex
\pdfoutput=1
\documentclass[12pt]{report}

\include{text/setups/setups}

\usepackage{caption}
\usepackage[english]{babel}
\usepackage{lineno, blindtext}
\usepackage{enumitem}
\usepackage{setspace}
\usepackage{float}
\floatstyle{plaintop}
\restylefloat{table}
\usepackage{bm}
\usepackage{booktabs}
\usepackage{array}
\usepackage{footnote}
\usepackage{afterpage}
\makesavenoteenv{tabular}
\makesavenoteenv{table}
\usepackage{footnotebackref}
\usepackage{amsmath}
\usepackage{colortbl}
\usepackage{makecell}
\usepackage{boldline}
\usepackage[toc,page]{appendix}
\usepackage{placeins}
\usepackage{chngcntr}
\newcolumntype{?}{!{\vrule width 1.5pt}}

\usepackage{multirow, hhline}
\usepackage{tablefootnote}

\usepackage{background}
\usepackage{lipsum}

\SetBgContents{CLAS12 Note 2021-002}
\SetBgScale{1.35}
\SetBgAngle{0}
\SetBgOpacity{1}
\SetBgColor{black}
\SetBgPosition{current page.south west}
\SetBgHshift{12cm}
\SetBgVshift{20cm}

\begin{document}

 \makeatletter
\newenvironment{sqcases}{%
  \matrix@check\sqcases\env@sqcases
}{%
  \endarray\right.%
}
\def\env@sqcases{%
  \let\@ifnextchar\new@ifnextchar
  \left\lbrack
  \def\arraystretch{1.2}%
  \array{@{}l@{\quad}l@{}}%
}
\makeatother

\newcommand\footnoteref[1]{\protected@xdef\@thefnmark{\ref{#1}}\@footnotemark}

\pdfcompresslevel=9
\pdfobjcompresslevel=9
\include{text/aliases/aliases}

\include{text/title_page/title}

\pagebreak


\include{text/toc/toc}

\include{text/introduction/intro}

\include{text/mm_distr/mm_distr}

\include{text/rad_eff/rad_eff}

\include{text/other_ch/other_ch}

\include{text/resolution/resolution}

\include{text/fermi/fermi}

\include{text/fsi/fsi}

\include{text/concl/concl}

\pagebreak
\newpage

\singlespacing
\bibliographystyle{ieeetr}
\bibliographystyle{apsrev4-1long}
\bibliography{note}

\end{document}

%% file: text/setups/setups.tex
\usepackage[utf8]{inputenc}
\usepackage{rotating}
\usepackage[usenames]{color}

\usepackage{graphicx}
\usepackage{grffile}

\usepackage{overpic}
\usepackage{url}
\usepackage{pst-plot}
\usepackage{amsmath,fullpage,graphicx,caption,float}
\usepackage{amssymb}
\usepackage{array,ragged2e,pst-node,pst-dbicons}
\usepackage{pstricks}
\usepackage{pst-node}
\usepackage{pst-blur}
\usepackage{comment}
\usepackage{color}
\usepackage{subfigure}
\usepackage{lscape}
\usepackage{multirow}
\usepackage{mathtools}
\usepackage{framed}
\usepackage{diagbox}

\usepackage{float}
\restylefloat{table}
\usepackage[margin=25mm]{geometry}

\renewcommand\baselinestretch{1.3}
\renewcommand\arraystretch{1.5}

\usepackage[numbers,sort&compress]{natbib}
\usepackage[nottoc]{tocbibind}

%% file: text/aliases/aliases.tex
 \def\be{\begin{eqnarray}}
 \def\ee{\end{eqnarray}}
 \def\ds{\displaystyle}

\newcommand{\mmm}{\mbox{íÜ÷}}
\newcommand{\reff}[1]{(\ref{#1})}
\newcommand{\ra}{\rangle}
\newcommand{\la}{\langle}
\newcommand{\rf}{\}}
\newcommand{\lf}{\{}
\newcommand{\ket}[1]{ | #1 \rangle }

%% file: text/title_page/title.tex
\noindent\begin{minipage}{\textwidth}
\begin{center}
\thispagestyle{empty}
\vspace{0.5cm}
{ \Large{Peculiar features of missing mass distributions in studies of exclusive reactions}}\\
\vspace{1cm}

{\large Iu.A. Skorodumina$^{1, a}$, G.V. Fedotov$^{2}$,  R.W. Gothe$^{1}$} \\[16pt]

\parbox{.86\textwidth}{\centering\footnotesize\it
$^1$Department of Physics and Astronomy, University of South Carolina, Columbia, SC\\[8pt]
\setstretch{0.3} 
$^2$National Research Centre ``Kurchatov Institute" B. P. Konstantinov Petersburg Nuclear Physics Institute, Gatchina, St. Petersburg, Russia\\
[20pt]
E-mail: $^a$skorodum@jlab.org}\\

\vspace{2cm}
{\bf Abstract}\\[9pt]

\end{center}
\everypar{\looseness=-1}
{\small This study examines the influence of the following five factors on missing mass distributions, radiative effects, admixtures from other channels, detector resolution, Fermi smearing, and final state interactions with spectator nucleons. All factors are considered independently of each other. The examination is supported by theoretical calculations and exemplified by distributions of the quantities $M^{2}_{X[0]}$ and $M^{2}_{X[\pi^{-}]}$ that are produced by the Monte Carlo simulation for the double-pion electroproduction off protons. The study also includes a naive modeling of kinematic effects of final state interactions with spectator nucleons.}

\end{minipage}

%% file: text/toc/toc.tex
\newpage
\renewcommand{\baselinestretch}{1}\normalsize
\setcounter{tocdepth}{3}
\setcounter{secnumdepth}{4}
\tableofcontents

\setcounter{page}{2}

%% file: text/introduction/intro.tex
\newpage
\chapter{Introduction}
\mbox{}\vspace{-\baselineskip}

Nowadays experimental investigations of exclusive meson photo- and electroproduction off protons have been put into high gear worldwide. When performing such an investigation, it is very important to properly isolate events that correspond to a particular reaction channel from the bulk of experimental data. This challenging task is commonly resolved employing the ``missing mass" technique.

During the data analysis, events of a desired exclusive channel are selected from the experimentally available event sample, which typically corresponds to the process $ep \rightarrow e'HX$, where $H$ denotes all registered final hadrons, while $X$ stands for all unregistered particles\footnote[1]{The scattered electron $e'$ must be registered here as it defines the reaction event.}. Then, a missing mass $M_{X}$ is a quantity that is calculated via energy-momentum conservation from the four-momenta of the registered particles ($e'$, $H$) and its distribution reflects the mass spectrum of the unregistered part ($X$). The examination of missing mass distributions then gives an opportunity to qualify the presence of any type of background in the investigated event sample and to judge the reliability of the entire event selection.

Although in general a missing mass of any number of unregistered particles can be considered, usually close attention is paid to the following two quantities, the missing mass squared of one missing hadron $h$ ($M^{2}_{X[h]}$) and the missing mass squared of the fully exclusive reaction ($M^{2}_{X[0]}$). These quantities serve the purpose of isolating the exclusive events best, as they form a discrete spectrum in the absence of such additional factors as detector resolution, radiative effects, background admixtures, etc. 

Employing the missing mass technique in the analysis allows using event samples with one unregistered final hadron along with the fully exclusive event sample, distinguishing in this way between so-called reaction ``topologies", each of which corresponds to a particular combination of registered final hadrons.  The number of available reaction topologies is then equal to $n+1$, where $n$ is the number of hadrons in the reaction final state. This approach allows an increase of the analyzed statistics (which is sometimes significant).

In order to pick out the exclusive reaction, in each topology the missing mass distribution is subject to a so-called ``exclusivity cut'' as a final step of the event selection. A properly chosen position of the exclusivity cut allows at least the suppression of the background and non-physical admixtures or even their complete removal.

Meanwhile, experimentally obtained missing mass distributions are inevitably folded with the influence of various factors such as detector resolution, radiative effects, background admixtures, etc. These factors disturb the missing mass distribution by changing its shape, width, and maximum position, hence complicating the task of isolating the desired exclusive channel. Therefore, to better understand the combined influence of these factors, it is extremely important to understand the impact of each factor by itself.

\everypar{\looseness=-1}
This study examines the influence of the following five factors on the missing mass distributions, i.e. radiative effects, admixture from other channels, detector resolution, Fermi smearing, and final state interactions with spectator nucleons. All factors are considered independently of each other. The examination is supported by theoretical calculations and exemplified by distributions of the quantities $M^{2}_{X[\pi^{-}]}$ and $M^{2}_{X[0]}$ that are produced by the Monte Carlo simulation for the reaction of double-pion electroproduction off protons. All conclusions, however, can be simply generalized for any missing particle and any exclusive reaction.

The Monte-Carlo simulation, which was performed to visually illustrate the manifestation of each factor, employs a particular well-established technique (see details in Sect.~3.1-3.4) for all considered factors except the last one, i.e. final state interactions with spectator nucleons. Unfortunately, no methods currently exist to simulate the latter. Therefore, to trace the impact of this factor on the missing quantities, a naive modeling of kinematic features of final state interactions with spectator nucleons is attempted in Sect.~\ref{sec:fsi}.

All histograms in this note are plotted under the following conditions, $E_{beam}$ = 2.039~GeV, 1.4~GeV $< W <$ 1.8~GeV and 0.4~GeV$^{2}$ $< Q^{2} <$ 0.6~GeV$^{2}$, and (unless specified otherwise) filled with events generated by the TWOPEG~\cite{twopeg} event generator. All distributions are normalized in a way that the maxima of the main peaks are equal to one.

To avoid confusion, the reader is strongly encouraged to pay close attention to the definition of the examined missing quantities, which is given in Chapter~\ref{ch:mm}, before considering any conclusions made in this study.

%% file: text/mm_distr/mm_distr.tex
\newpage
\chapter{Missing mass distributions}
\label{ch:mm}
\mbox{}\vspace{-\baselineskip}

Let's consider a sample of events that correspond to the reaction of double-pion electroproduction off protons $ep\rightarrow e'p'\pi^{+}\pi^{-}$. Then the following missing quantities can be defined,
\begin{equation}
\begin{aligned}
&M_{X[0]}^{2}&=&\left [P_{X[0]}^{\mu} \right ]^{2}&=&~\left (P_{e}^{\mu} + P_{p}^{\mu}- P_{e'}^{\mu}- P_{p'}^{\mu}-  P_{\pi^{+}}^{\mu} - P_{\pi^{-}}^{\mu}\right )^{2}~~\textrm{and}\\
&M_{X[\pi^{-}]}^{2}&=&\left [P_{X[\pi^{-}]}^{\mu}\right ]^{2}&=&~\left (P_{e}^{\mu} + P_{p}^{\mu}- P_{e'}^{\mu}- P_{p'}^{\mu}-  P_{\pi^{+}}^{\mu}\right )^{2},\\
\end{aligned}\label{eq:mm_def}
\end{equation}
where $P_{X[0]}^{\mu}$ and $P_{X[\pi^{-}]}^{\mu}$ are the corresponding missing four-vectors, while $P_{i}^{\mu}$ is the four-momentum of the particle $i$.

Let's first assume that (i) the target proton is free and at rest, (ii) no events from other channels are present in the sample, (iii) all four-momenta $P_{i}^{\mu}$ are defined exactly without any uncertainty, and (iv) neither radiative effects nor FSI occur. 

\begin{figure}[htp]
\begin{center}
\framebox{\includegraphics[width=\textwidth]{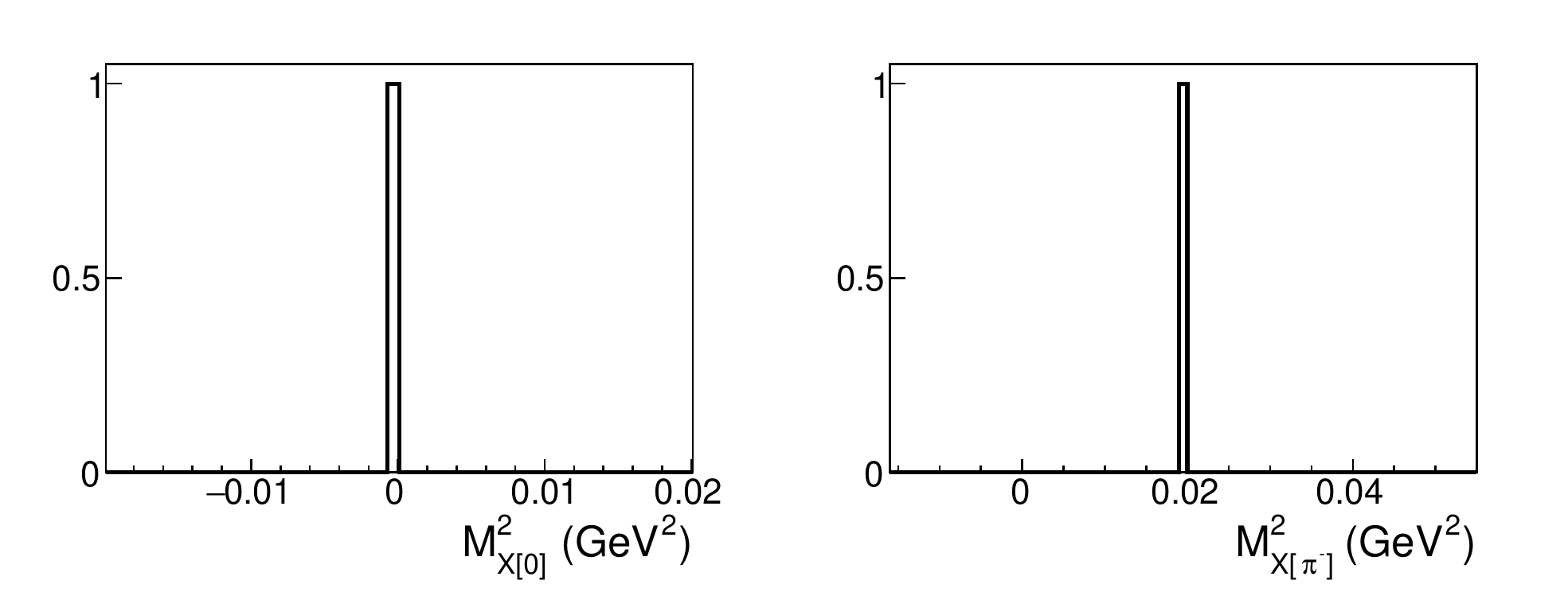}}
\caption{\small Simulated distributions of the quantities $M_{X[0]}^{2}$ (left) and $M_{X[\pi^{-}]}^{2}$ (right) plotted under the assumptions specified in the text.} \label{fig:norad_nofsi}
\end{center}
\end{figure}

Then, using energy-momentum conservation and the fact that the particles are on their mass shell, the following relations can be obtained,
\begin{equation}
\begin{aligned}
&M_{X[0]}^{2}&=&\left [P_{X[0]}^{\mu} \right ]^{2}&=&~0~~~~\textrm{and}\\
&M_{X[\pi^{-}]}^{2}&=&\left [P_{X[\pi^{-}]}^{\mu}\right ]^{2}&=&~[P_{\pi^{-}}^{\mu}]^{2}=m_{\pi^{-}}^{2},\\
\end{aligned}\label{eq:plain}
\end{equation}
where $m_{\pi^{-}}$ is the mass of the negative pion.

As follows from Eqs.~\eqref{eq:plain}, both $M_{X[0]}^{2}$ and $M_{X[\pi^{-}]}^{2}$ form a discrete narrow peak at the position of zero and $m_{\pi^{-}}^{2}$, respectively. This is illustrated in Fig.~\ref{fig:norad_nofsi}, where the simulated distributions of both missing quantities are shown. Note that for the quantity $M_{X[0]}^{2}$ the missing four-vector $P_{X[0]}^{\mu}$ in Eqs.~\eqref{eq:plain} is equal to zero componentwise, which means that the energy and each momentum components are equal to zero.

In the next Chapter, the influence of different factors on the distributions of $M_{X[0]}^{2}$ and $M_{X[\pi^{-}]}^{2}$ is examined. All factors are considered independently.

%% file: text/rad_eff/rad_eff.tex
\newpage
\chapter{Influence of different factors on missing mass distributions}
\section{Radiative effects}
\mbox{}\vspace{-\baselineskip}

Let's now assume that for some events in the sample either the incoming or scattered electron undergoes a radiative photon emission, and hence changes its energy either before or after the reaction, respectively\footnote[2]{If the incoming electron emits, the actual reaction beam energy turns out to be lower than the nominal value. If the final electron emits, its registered energy turns out to be lower than its actual value.}. 

For experimentally collected events, the information about such emissions (and the corresponding changes in electron kinematics) is typically not accessible. As a consequence, the missing quantities are calculated using the four-momentum of the incoming electron $P_{e}^{\mu}$ determined before the emission and the four-momentum of the scattered electron $P_{e'}^{\mu}$ determined after the emission.

Therefore, for the events affected by radiative effects, the missing quantities $M_{X[0]}^{2}$ and $M_{X[\pi^{-}]}^{2}$ can be expressed in the following way,\vspace{-0.125em}
\begin{equation}
\begin{aligned}
&M_{X[0]}^{2}&=&~\left [P_{X[0]}^{\mu} \right ]^{2}&=&~[P^{\mu}_{\gamma}]^{2}=0~~\textrm{and}\\[8pt]
&M_{X[\pi^{-}]}^{2}&=&\left [P_{X[\pi^{-}]}^{\mu}\right ]^{2}&=&~(P_{\pi^{-}}^{\mu}+P^{\mu}_{\gamma})^{2}=[P^{\mu}_{\pi^{-}}]^{2} +[P^{\mu}_{\gamma}]^{2}+2\left [P_{\pi^{-}}\right ]_{\mu} P_{\gamma}^{\mu} \\
&&&&=&~m_{\pi^{-}}^{2} +2(E_{\pi^{-}}E_{\gamma} - (\overrightarrow{p}_{\pi^{-}}\cdot \overrightarrow{p}_{\gamma}))  \\
&&&&=&~m_{\pi^{-}}^{2} +2(E_{\pi^{-}}E_{\gamma} - |\overrightarrow{p}_{\pi^{-}}|E_{\gamma}\cos\beta) \\
&&&&=&~ m_{\pi^{-}}^{2} +2E_{\gamma}(E_{\pi^{-}} -|\overrightarrow{p}_{\pi^{-}}|\cos\beta ) >m_{\pi^{-}}^{2}   ,\\
\end{aligned}\label{eq:mm_rad}
\end{equation}
where $P^{\mu}_{\gamma}$ is the four-momentum of the emitted radiative photon, $E_{i}$ and $\overrightarrow{p}_{i}$ are the energy and the three-momentum of the particle $i$, respectively, while $\beta$ corresponds to the angle between the $\pi^{-}$ and the photon.

\begin{figure}[htp]
\begin{center}
\framebox{\includegraphics[width=\textwidth]{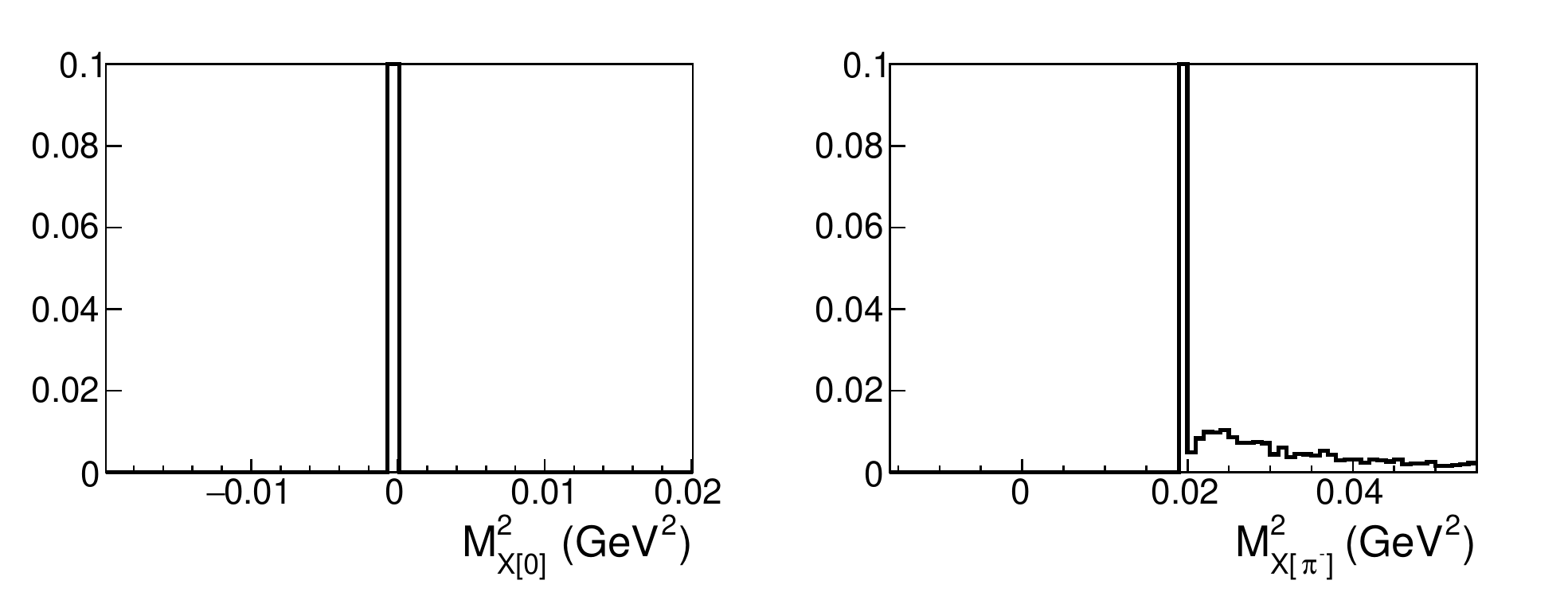}}
\caption{\small Impact of radiative effects on $M_{X[0]}^{2}$ (left) and $M_{X[\pi^{-}]}^{2}$ (right). Both distributions are zoomed in on small $y$ to make the impact of radiative effects visible.} \label{fig:mm_rad}
\end{center}
\end{figure}

As follows from Eqs.~\eqref{eq:mm_rad}, the quantity $M_{X[0]}^{2}$ feels no impact of radiative photon emissions as radiated events turn out to contribute to the main distribution peak at zero. Meanwhile, in the distribution of $M_{X[\pi^{-}]}^{2}$, the main peak at $m_{\pi^{-}}^{2}$ acquires a right-side tail, populated with events with photon emissions. This situation is illustrated in Fig.~\ref{fig:mm_rad}, which shows the distributions of $M_{X[0]}^{2}$ (left) and $M_{X[\pi^{-}]}^{2}$ (right) plotted for the event sample affected by radiative effects\footnote[3]{The TWOPEG event generator~\cite{twopeg}, which was used to produce these histograms, employs radiative effects according to the well-known approach from Ref.~\cite{Mo:1968cg}. The minimal photon energy was set to 10~MeV.}.

Note that for the quantity $M_{X[0]}^{2}$ the missing four-vector $P_{X[0]}^{\mu}$ in Eqs.~\eqref{eq:mm_rad} turns out to be the four-momentum of the radiative photon, which is not equal to zero componentwise. However, being massless, the photon has the energy equal to its momentum magnitude, which gives zero upon the calculation of $M_{X[0]}^{2}$. Thus, the zero value of $M_{X[0]}^{2}$ has a different nature for events with and without radiative effects.

%% file: text/other_ch/other_ch.tex
\section{Admixture from other channels}
\mbox{}\vspace{-\baselineskip}

Let's assume that some events in the sample correspond to the background channel with a greater number of final state particles, i.e. $ep\rightarrow e'p'\pi^{+}\pi^{-}x$. Then, for such background events, the following relations for the missing quantities can be obtained,
\begin{equation}
\begin{aligned}
&M_{X[0]}^{2}&=&~\left [P_{X[0]}^{\mu} \right ]^{2}&=&~[P^{\mu}_{x}]^{2} = m_{x}^{2} >0~~\textrm{and}\\
&M_{X[\pi^{-}]}^{2}&=&\left [P_{X[\pi^{-}]}^{\mu}\right ]^{2}&=&~(P_{\pi^{-}}^{\mu}+P^{\mu}_{x})^{2}=[P^{\mu}_{\pi^{-}}]^{2} +[P^{\mu}_{x}]^{2}+2\left [P_{\pi^{-}}\right ]_{\mu} P_{x}^{\mu} \\
&&&&=&~m_{\pi^{-}}^{2} + m_{x}^{2} +2(E_{\pi^{-}}E_{x} - (\overrightarrow{p}_{\pi^{-}}\cdot \overrightarrow{p}_{x}))  \\
&&&&>&~m_{\pi^{-}}^{2} + m_{x}^{2}+2m_{\pi^{-}}m_{x} > m_{\pi^{-}}^{2},\\
\end{aligned}\label{eq:mm_other_ch}
\end{equation}
where $P^{\mu}_{x}$ is the four-momentum of the extra particle $x$ and $m_{x}$ is its mass.

Thus, in the distribution of the quantity $M_{X[0]}^{2}$, the background events form an additional discrete narrow peak located on the right side of the main peak at the position of $m_{x}^{2}$. Meanwhile, the background in the distribution of $M_{X[\pi^{-}]}^{2}$, being also well-separated from the main peak and located on its right side, forms a broad structure, which starts at the value of $m_{\pi^{-}}^{2} + m_{x}^{2}+2m_{\pi^{-}}m_{x}$.

\begin{figure}[htp]
\begin{center}
\framebox{\includegraphics[width=\textwidth]{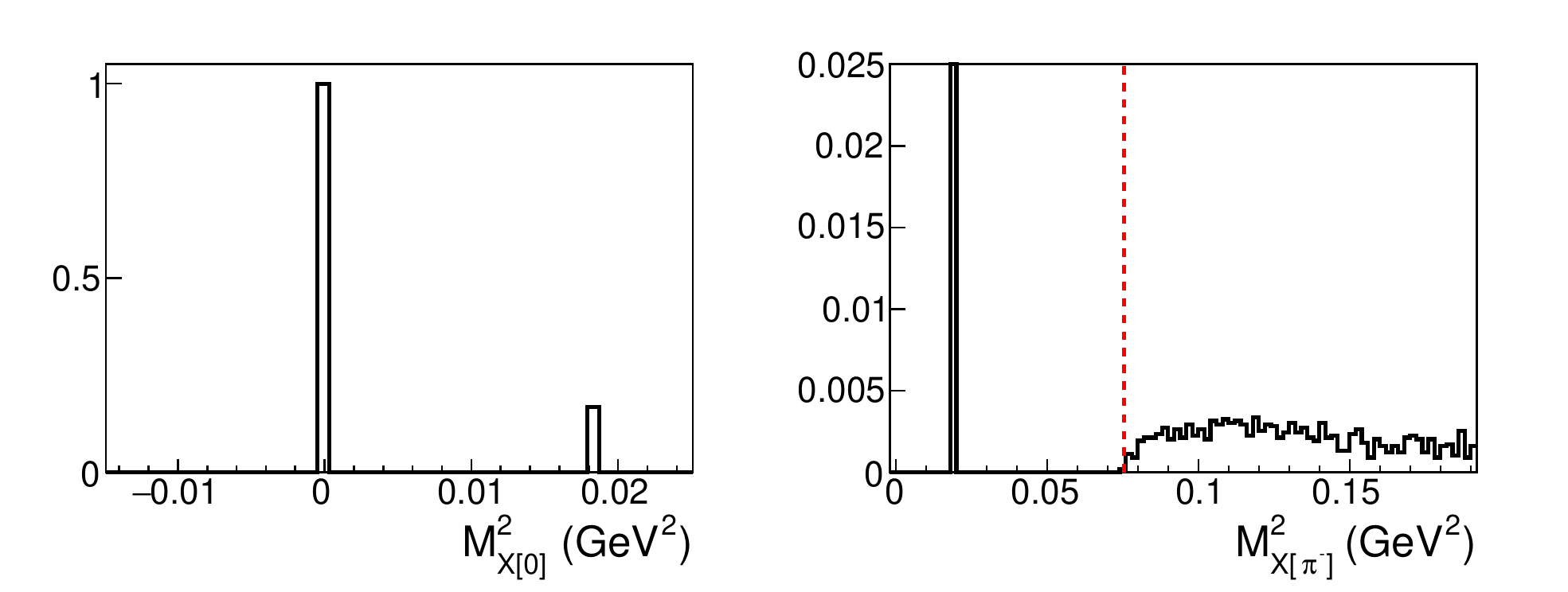}}
\caption{\small Distributions of the quantities $M_{X[0]}^{2}$ (left) and $M_{X[\pi^{-}]}^{2}$ (right) plotted for the event sample that contains the background admixture from the channel $ep\rightarrow e'p'\pi^{+}\pi^{-}\pi^{0}$. The discrete right peak (left panel) and the right-side structure (right panel), both well-separated from the main peak, are formed by the background events. The right plot is zoomed in on small $y$, and the red dashed line there marks the value of $m_{\pi^{-}}^{2} + m_{\pi^{0}}^{2}+2m_{\pi^{-}}m_{\pi^{0}}$.   } \label{fig:mm_backgr}
\end{center}
\end{figure}

This situation is illustrated in Fig.~\ref{fig:mm_backgr} for the case when the extra particle is $\pi^{0}$. The plots are produced by means of the GENEV event generator~\cite{Genev}.

%% file: text/resolution/resolution.tex
\section{Detector resolution}
\mbox{}\vspace{-\baselineskip}

Let's now assume that for all events in the sample the momenta of registered particles are determined within the uncertainty due to the detector resolution. Then the missing quantities $M_{X[0]}^{2}$ and $M_{X[\pi^{-}]}^{2}$ acquire some uncertainty as well, which is estimated below.

A missing quantity $M_{X}^{2}$ can be written in the following way,
\begin{equation}
\begin{aligned}
&M_{X}^{2}&=&~(E_{X})^{2} - (p_{X}^{x})^{2} - (p_{X}^{y})^{2}  -  (p_{X}^{z})^{2} \\
&&=&~\left (\sum_{\substack{i}} \pm \sqrt{m_{i}^{2}+p_{i}^{2}} \right )^{2} - \left (\sum_{\substack{i}}\pm p_{i}^{x} \right )^{2} - \left (\sum_{\substack{i}}\pm p_{i}^{y} \right )^{2} - \left (\sum_{\substack{i}}\pm p_{i}^{z} \right )^{2},\\
\end{aligned}\label{eq:mm_def2}
\end{equation}
where $E_{X}$ and $p_{X}^{j}$ ($j = x,~y,~z$) are the energy and momentum components of the missing four-vector, while $m_{i}$, $E_{i}$, and $p_{i}^{j}$ are the mass, energy, and momentum components of the individual particles with the index $i$ running over all particles involved in the missing mass calculation (see Eqs.~\eqref{eq:mm_def}). For the $\pm$ sign, the plus is taken for the initial particles ($e$ and $p$) and the minus for the final particles.

First, one can estimate the uncertainties of the quantities $E_{X}$ and $p_{X}^{j}$ through the momentum uncertainties of the individual particles. 
The absolute uncertainties for the energy components $E_{X}$ can be expressed by
\begin{equation}
\begin{aligned}
&\Delta E_{X[0]} &=&~\sqrt{ \left ( \Delta E_{e'} \right )^{2} + \left ( \Delta E_{p'} \right )^{2} +  \left ( \Delta E_{\pi^{+}} \right )^{2} + \left ( \Delta E_{\pi^{-}} \right )^{2}} \\
&&=&~\sqrt{\left ( \Delta p_{e'} \right )^{2} + \left (p_{p'}/E_{p'}\right )^{2} \left (\Delta p_{p'} \right )^{2} +  \left (p_{\pi^{+}}/E_{\pi^{+}}\right )^{2} \left (\Delta p_{\pi^{+}} \right )^{2}+\left (p_{\pi^{-}}/E_{\pi^{-}}\right )^{2} \left (\Delta p_{\pi^{-}} \right )^{2}}, \\[10pt]
&\Delta E_{X[\pi^{-}]} &=&~\sqrt{ \left ( \Delta E_{e'} \right )^{2} + \left ( \Delta E_{p'} \right )^{2} + \left ( \Delta E_{\pi^{+}} \right )^{2}}\\
&&=&~\sqrt{\left ( \Delta p_{e'} \right )^{2} + \left (p_{p'}/E_{p'}\right )^{2} \left (\Delta p_{p'} \right )^{2} +  \left (p_{\pi^{+}}/E_{\pi^{+}}\right )^{2} \left (\Delta p_{\pi^{+}} \right )^{2}},\\
\end{aligned}\label{eq:res}
\end{equation}
where $\Delta p_{i}$ is the uncertainty of the momentum magnitude for the particle $i$, which comes from the track momentum resolution of the Drift Chambers.

The absolute uncertainties for the momentum components $p_{X}^{j}$ are in turn given by
\begin{equation}
\begin{aligned}
&\Delta p_{X[0]}^{j} &=&~\sqrt{ \left ( \Delta p_{e'}^{j} \right )^{2} + \left ( \Delta p_{p'}^{j} \right )^{2} +  \left ( \Delta p_{\pi^{+}}^{j} \right )^{2} + \left ( \Delta p_{\pi^{-}}^{j} \right )^{2}}, \\
&\Delta p_{X[\pi^{-}]}^{j} &=&~\sqrt{ \left ( \Delta p_{e'}^{j} \right )^{2} + \left ( \Delta p_{p'}^{j} \right )^{2} +  \left ( \Delta p_{\pi^{+}}^{j} \right )^{2}},\\
\end{aligned}\label{eq:res2}
\end{equation}
where $\Delta p_{i}^{j}$ are the uncertainties of the components of the particle's momenta ($j~=~x,~y,~z$), which come from the track momentum and angular resolutions of the Drift~Chambers.

As follows from Eqs.~\eqref{eq:res} and~\eqref{eq:res2}, the absolute uncertainties of $E_{X[0]}$ and $p_{X[0]}^{j}$ are larger than those of $E_{X[\pi^{-}]}$ and $p_{X[\pi^{-}]}^{j}$, respectively, as the former quantities include extra terms associated with the additional registered particle (which is the $\pi^{-}$ here).

Now, using Eq.~\eqref{eq:mm_def2}, the absolute uncertainties of the missing quantities $M_{X[0]}^{2}$ and $M_{X[\pi^{-}]}^{2}$ can be estimated as the following,
\begin{equation}
\begin{aligned}
&\Delta M_{X[0]}^{2} &=&~\sqrt{ \left (2E_{X[0]} \Delta E_{X[0]} \right )^{2} + \sum_{\substack{j = x,~y,~z}}\left (2p_{X[0]}^{j} \Delta p_{X[0]}^{j} \right )^{2}}, \\
&\Delta M_{X[\pi^{-}]}^{2} &=&~\sqrt{ \left (2E_{X[\pi^{-}]} \Delta E_{X[\pi^{-}]} \right )^{2} +  \sum_{\substack{j = x,~y,~z}}\left (2p_{X[\pi^{-}]}^{j} \Delta p_{X[\pi^{-}]}^{j} \right )^{2}}. \\
\end{aligned}\label{eq:res3}
\end{equation}

In Eqs.~\eqref{eq:res3} the uncertainties $\Delta E_{X[0]}$, $\Delta p_{X[0]}^{j}$ and $ \Delta E_{X[\pi^{-}]}$, $\Delta p_{X[\pi^{-}]}^{j}$ are respectively comparable, though the former are systematically larger than the latter (as was shown above). Meanwhile, both $E_{X[0]}$ and $p_{X[0]}^{j}$ are very close to zero, while both $E_{X[\pi^{-}]}$ and $p^{j}_{X[\pi^{-}]}$ are non-zero. As a consequence, the quantity $M_{X[0]}^{2}$ acquires smaller absolute uncertainty value than $M_{X[\pi^{-}]}^{2} $. This is, however, not the case for their relative uncertainties because (in contrast to $M_{X[\pi^{-}]}^{2}$) the quantity $M_{X[0]}^{2}$ is extremely close to zero.

\begin{figure}[htp]
\begin{center}
\framebox{\includegraphics[width=\textwidth]{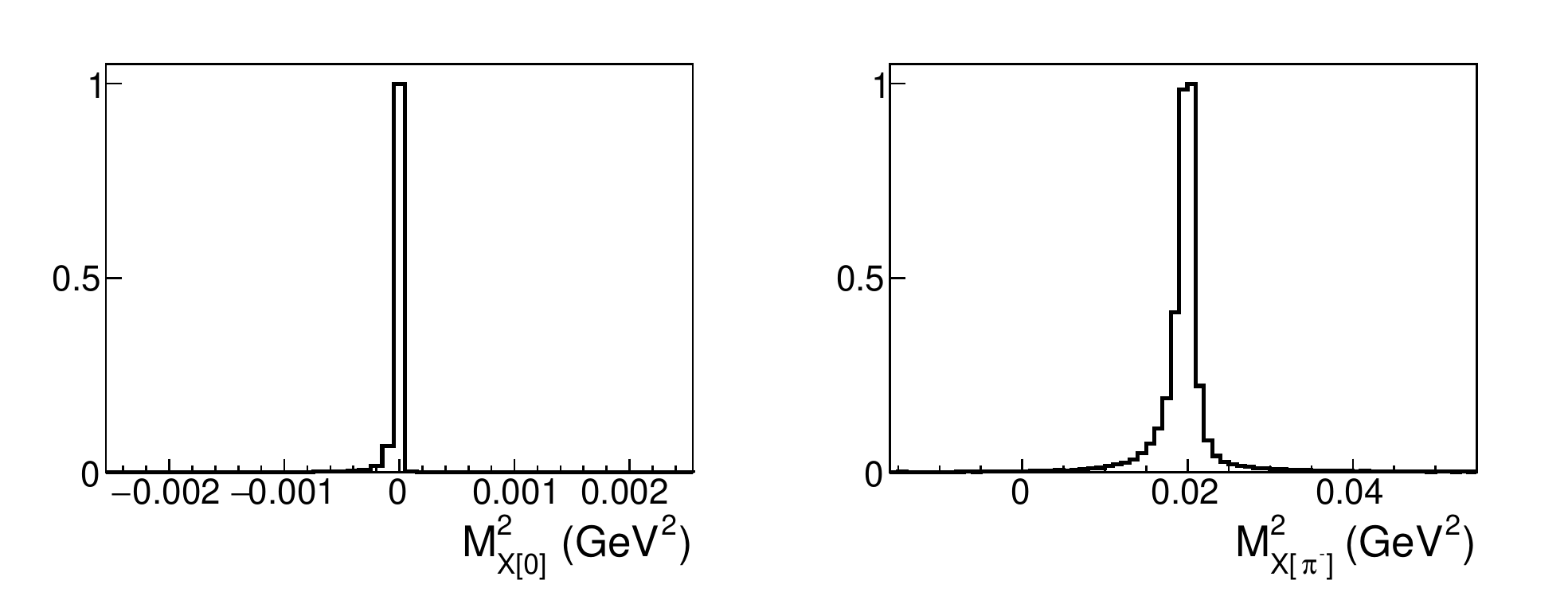}}
\caption{\small Impact of the detector resolution on $M_{X[0]}^{2}$ (left) and $M_{X[\pi^{-}]}^{2}$ (right). The distribution of $M_{X[0]}^{2}$ is zoomed in on $x$ to demonstrate the disturbances. To produce these histograms, events generated with TWOPEG~\cite{twopeg} were reconstructed via the CLAS reconstruction software~\cite{Mecking:2003zu}.} \label{fig:mm_res}
\end{center}
\end{figure}

To simulate the impact of the detector resolution on the missing quantities, events generated with TWOPEG~\cite{twopeg} were reconstructed via the CLAS reconstruction software~\cite{Mecking:2003zu}. Figure~\ref{fig:mm_res} presents the resulting distributions, which illustrate the above calculations. As seen in the plots, the $M_{X[0]}^{2}$ distribution (left) turns out to be visually very narrow with slight disturbances, while the $M_{X[\pi^{-}]}^{2}$ distributions (right) acquires perceptible smearing\footnote[4]{Note that the detector resolution depends on particle kinematics~\cite{Mecking:2003zu}.}.

%% file: text/fermi/fermi.tex
\section{Fermi smearing}
\mbox{}\vspace{-\baselineskip}

Let's now assume that for all events in the sample the reaction happens off a moving proton, as in the case when the proton is contained within a nucleus. Then the four-momentum of the target proton has the following components, 
\begin{equation}
\begin{aligned}
P_{p}^{\mu} = (\sqrt{m^{2}_{p}+p^{2}_{F}},~p_{F}^{x},~p_{F}^{y},~p_{F}^{z}),
\end{aligned}\label{eq:p_4mom}
\end{equation}
where $p_{F}^{j}$ ($j=x,~y,~z$) are the components of the proton three-momentum (or the Fermi-momentum) and $p_{F}$ is its magnitude. Note that the proton is assumed to be on mass-shell.

If for each event in the sample the components of the Fermi momentum are exactly known without any uncertainty, then $P_{p}^{\mu}$ given by Eq.~\eqref{eq:p_4mom} can be used to calculate the missing quantities $M_{X[0]}^{2}$ and $M_{X[\pi^{-}]}^{2}$. Such a calculation will then result in Eqs.~\eqref{eq:plain}, which correspond to the illustration in Fig.~\ref{fig:norad_nofsi}.

However, for event samples collected experimentally, the information on the initial proton momentum is typically not accessible or incomplete, which leads to the inevitability to conduct the analysis under the target-at-rest assumption. This means that the following form of the initial proton four-momentum is used to calculate the missing quantities,
\begin{equation}
\begin{aligned}
\widetilde{P}_{p}^{\mu} = (m_{p},~0,~0,~0).
\end{aligned}\label{eq:p_4mom_rest}
\end{equation}

Now one can proceed to estimating the quantities $M_{X[0]}^{2}$ and $M_{X[\pi^{-}]}^{2}$ under the target-at-rest assumption. The former can be expressed by
\begin{equation}
\begin{aligned}
&M_{X[0]}^{2}&=&~\left [P_{X[0]}^{\mu} \right ]^{2}&=&~ \left (\widetilde{P}^{\mu}_{p} - P^{\mu}_{p} \right )^{2}=[\widetilde{P}^{\mu}_{p} ]^{2} + [P^{\mu}_{p}]^{2}-2 [\widetilde{P}_{p} ]_{\mu} P_{p}^{\mu}\\
&&&&=&~2m_{p}^{2}-2m_{p}\sqrt{m_{p}^{2}+p_{F}^{2}}=2m_{p}\left (m_{p} - \sqrt{m_{p}^{2}+p_{F}^{2}} \right ) < 0.
\end{aligned}\label{eq:mm0_fermi}
\end{equation}

The final expression in Eq.~\eqref{eq:mm0_fermi} depends on the Fermi momentum magnitude (but not on its direction) and is independent on the reaction kinematics. The $M_{X[0]}^{2}$ distribution thus acquires the left-sided smearing and kinematically stable shape.


Meanwhile, the quantity $M_{X[\pi^{-}]}^{2}$ can be written in the following way,
\begin{equation}
\begin{aligned}
&M_{X[\pi^{-}]}^{2}&=&~\left [P_{X[\pi^{-}]}^{\mu} \right ]^{2}&=&~\left (P_{\pi^{-}}^{\mu}+ \widetilde{P}^{\mu}_{p} - P^{\mu}_{p} \right )^{2}\\
&&&& =&~\left [P^{\mu}_{\pi^{-}} \right ]^{2} + \left (  \widetilde{P}^{\mu}_{p} - P^{\mu}_{p} \right )^{2}+2[P_{\pi^{-}}]_{\mu} \left ( \widetilde{P}^{\mu}_{p}  -  P^{\mu}_{p}  \right )\\
&&&&=&~m_{\pi^{-}}^{2} + M_{X[0]}^{2} +2\left \{ E_{\pi^{-}}\left (m_{p}-\sqrt{m_{p}^{2}+p_{F}^{2}}\right ) + (\overrightarrow{p}_{\pi^{-}} \cdot \overrightarrow{p}_{F}) \right \}\\
&&&&=&~m_{\pi^{-}}^{2} + 2\left (m_{p}+E_{\pi^{-}}\right )\left (m_{p}-\sqrt{m_{p}^{2}+p_{F}^{2}}\right ) +2\left (\overrightarrow{p}_{\pi^{-}} \cdot \overrightarrow{p}_{F} \right ),\\
\end{aligned}\label{eq:mmpim_fermi}
\end{equation}
where for $M_{X[0]}^{2}$ the expression obtained in Eq.~\eqref{eq:mm0_fermi} was employed.

The final expression in Eq.~\eqref{eq:mmpim_fermi} depends on the magnitude and direction of the Fermi momentum as well as the reaction kinematics. It is also noteworthy that the second summand in this expression is always negative, while the third summand can have either sign. Thus, the value of $M_{X[\pi^{-}]}^{2}$ can be either greater or smaller than $m_{\pi^{-}}^{2}$, depending on the interplay between the two terms. All these lead to both-sided smearing of the $M_{X[\pi^{-}]}^{2}$ distribution and kinematic dependence of its shape.

\begin{figure}[htp]
\begin{center}
\framebox{\includegraphics[width=\textwidth]{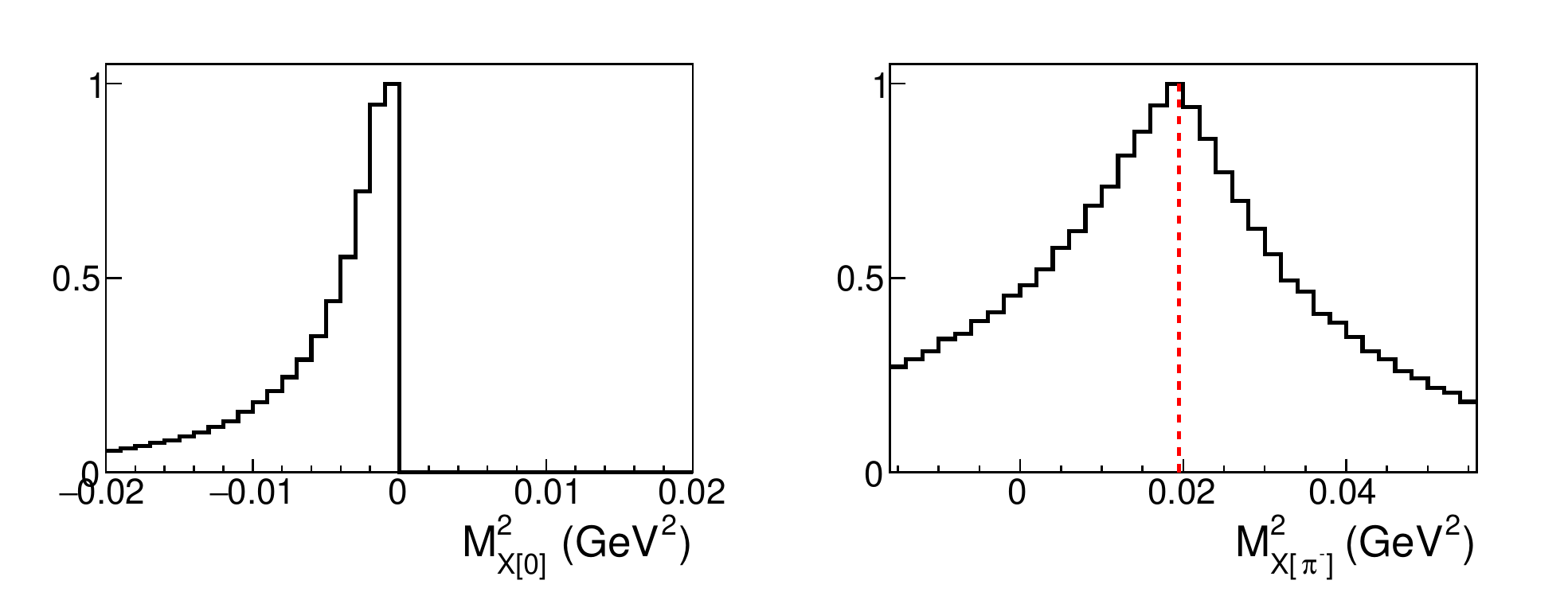}}
\caption{\small Quantities $M_{X[0]}^{2}$ (left) and $M_{X[\pi^{-}]}^{2}$ (right) affected by the Fermi smearing. The distributions are plotted under the target-at-rest assumption for the reaction occurring off protons that move in deuterium nuclei. To produce the plots, the TWOPEG-D event generator~\cite{twopeg-d} was used, which assumes the reaction to happen in the quasi-free regime and the momentum of the initial proton to be distributed according to the Bonn potential~\cite{Machleidt:1987hj}. The red dashed line in the right plot marks the value of $m_{\pi^{-}}^{2}$.} \label{fig:mm_fermi}
\end{center}
\end{figure}

The foregoing calculations are illustrated in Fig.~\ref{fig:mm_fermi}, which shows the distributions of the missing quantities plotted under the target-at-rest assumption for the reaction occurring off protons that move in deuterium nuclei. These distributions are produced by the TWOPEG-D event generator~\cite{twopeg-d}, which assumes the reaction to happen in the quasi-free regime and the momentum of the initial proton to be distributed according to the Bonn potential~\cite{Machleidt:1987hj}. As seen in the plots, distributions of both missing quantities demonstrate a considerable Fermi smearing, which is left-sided for $M_{X[0]}^{2}$ and both-sided for $M_{X[\pi^{-}]}^{2}$. 

The kinematic dependence of the $M_{X[\pi^{-}]}^{2}$ distribution is worth special mentioning. For $W$ close to the threshold, its shape is strongly asymmetric with the event excess at the left. As $W$ grows, the asymmetry gradually vanishes, and at $W\sim $ 1.7-1.8~GeV the distribution becomes symmetric. In addition to that, the distribution smearing increases with growing~$W$. The plot in the right panel of Fig.~\ref{fig:mm_fermi} is produced for the $W$ range from 1.4~GeV to 1.8~GeV, and hence demonstrates the combined impact of these effects.

%% file: text/fsi/fsi.tex
\section{Final state interactions with spectator nucleons}
\mbox{}\vspace{-\baselineskip}
\label{sec:fsi}

Let's again consider the case when the exclusive reaction happens off a proton that is contained within a nucleus. The final hadrons, once produced, can then experience interactions with spectator nucleons. Meanwhile, spectator nucleons are extrinsic to the original exclusive reaction, and therefore any interaction with them breaks the energy-momentum conservation imposed on the reaction particles. As a consequence, final state interactions (FSI) with spectator nucleons introduce disturbances to the distributions of missing quantities.

In this Section, the influence of such interactions on distributions of considered missing quantities is traced. This task, however, is complicated by 
the fact that no methods to properly simulate FSI effects currently exist due to their complex nature.

\everypar{\looseness=-1}
Therefore, in this Section, a naive modeling of FSI with spectator nucleons is attempted. The basic idea that underlies this modeling is that all mechanisms that possibly can happen during FSI have one simple feature in common; namely, they alter the momentum of the participating hadrons. 
This feature may be used to track manifestations of FSI effects kinematically. The modeling is hence of kinematic character, as it fully focuses on hadron momentum alterations and disregards particular mechanisms behind these alterations.

To perform such a modeling, one can assume that the final hadrons of the considered exclusive reaction experience momentum alterations as in the case of FSI with spectator nucleons\footnote[5]{In general this modeling suits for any kind of momentum alterations that break the energy-momentum conservation between the reaction particles.}. For simplicity, the following assumptions can be introduced, (i) only one registered final hadron is affected, (ii) the type of the affected hadron is the same among all affected events, and (iii) only the magnitude of the hadron momentum changes\footnote[6]{The adequacy of the latter assumption may vary depending on the type of the affected hadron, the type (gain/loss) and degree of momentum alterations, and the underlying interaction mechanism.}.

Then for each event, the alterations of the hadron momentum can be parameterized by $p'_{h} = \varepsilon p_{h}$, where $p_{h}$ and $p'_{h}$ are the momentum magnitudes of the hadron before and after the alteration, respectively, and $\varepsilon >0$ reflects the alteration degree. One should also keep in mind that the variable $\varepsilon$ may have some distribution among events in the sample, and hence it is convenient to introduce here $\xi(\varepsilon)$ as the corresponding probability density function, which will come into play later in this Section.

\everypar{\looseness=-1}
Now, in order to trace the influence of FSI with spectator nucleons on the missing quantities $M_{X[0]}^{2}$ and $M_{X[\pi^{-}]}^{2}$, one can estimate these quantities for events affected by the aforementioned momentum alterations, i.e. for those events with $\varepsilon \neq $1. The quantity $M_{X[0]}^{2}$ can be expressed by\vspace{-0.69em}
\begin{equation}
\begin{aligned}
&M_{X[0]}^{2}&=&~(P^{\mu}_{h} -P'^{\mu}_{h})^{2} = [P^{\mu}_{h}]^{2} +\left [P'^{\mu}_{h}\right ]^{2}-2[P_{h}]_{\mu} P'^{\mu}_{h} \\
&&=&~2m^{2}_{h}-2 \left (E_{h}E'_{h} - (\overrightarrow{p}_{h}\cdot \overrightarrow{p}'_{h}) \right )\\
&&=&~2m^{2}_{h} - 2\left (\sqrt{m^{2}_{h}+p^{2}_{h}}\sqrt{m^{2}_{h}+\varepsilon^{2}p^{2}_{h}}-\varepsilon p^{2}_{h}\right ),\\[-7pt]
\end{aligned}\label{eq:mm0_fsi}
\end{equation}
where $m_{h}$ is the mass of the hadron that undergoes the momentum change.

\newpage

The final expression in Eq.~\eqref{eq:mm0_fsi} is always less than zero, regardless of both the value of $\varepsilon$ and hadron kinematics, as the comparison below demonstrates\footnote[7]{In this comparison the lower indices are dropped and the wedge symbol ($\wedge$) means ``compare~with".}.
\vspace{-0.3em}
\begin{equation}
\begin{aligned}
2m^{2} - 2(\sqrt{m^{2}+p^{2}}\sqrt{m^{2}+\varepsilon^{2}p^{2}}-\varepsilon p^{2})& ~~\wedge~~ 0&\\
m^{2} - \sqrt{m^{2}+p^{2}}\sqrt{m^{2}+\varepsilon^{2}p^{2}} +\varepsilon p^{2} &~~\wedge~~ 0&\\
m^{2}+\varepsilon p^{2} &~~\wedge~~ \sqrt{m^{2}+p^{2}}\sqrt{m^{2}+\varepsilon^{2}p^{2}}&\\
m^{4}+\varepsilon^{2}p^{4}+2m^{2}\varepsilon p^{2} &~~\wedge~~ m^{4} + p^{2}m^{2} + m^{2}\varepsilon^{2}p^{2}+\varepsilon^{2}p^{4}&\\
2m^{2}\varepsilon p^{2} &~~\wedge~~ p^{2}m^{2}+  m^{2}\varepsilon^{2}p^{2}&\\
0 &~~\wedge~~ p^{2}m^{2}(\varepsilon^{2}-2\varepsilon+1)&\\
0&~~<~~ p^{2}m^{2}(\varepsilon-1)^{2}&\\[-8pt]
\end{aligned}\label{eq:comp}
\end{equation}
The quantity $M_{X[\pi^{-}]}^{2}$ in turn can be written as\vspace{-0.35em}
\begin{equation}
\begin{aligned}
&M_{X[\pi^{-}]}^{2}&=&~(P^{\mu}_{\pi^{-}}+ P^{\mu}_{h} -P'^{\mu}_{h})^{2} \\
&&=&~[P^{\mu}_{\pi^{-}}]^{2} +(P^{\mu}_{h} -P'^{\mu}_{h})^{2}+2\left [P_{\pi^{-}}\right ]_{\mu} \left (P^{\mu}_{h} -P'^{\mu}_{h}\right ) \\
&&=&~m_{\pi^{-}}^{2} + M_{X[0]}^{2} +2\left \{E_{\pi^{-}}(E_{h}-E'_{h}) - \overrightarrow{p}_{\pi^{-}}\cdot(\overrightarrow{p}_{h} - \overrightarrow{p}'_{h})\right \}\\
&&=&~m_{\pi^{-}}^{2} + M_{X[0]}^{2} +2\left \{(\overrightarrow{p}_{\pi^{-}}\cdot \overrightarrow{p}_{h})(\varepsilon-1) - E_{\pi^{-}}(E'_{h}-E_{h})\right \}.\\[-7pt]
\end{aligned}\label{eq:mm_pim_fsi}
\end{equation}

The final expression in Eq.~\eqref{eq:mm_pim_fsi} can be either greater or smaller than $m_{\pi^{-}}^{2}$ depending on the interrelation between the value of $M_{X[0]}^{2}$ (which is always negative) and the quantity in the curly brackets (which can have either sign).

\everypar{\looseness=-1}
As follows from Eqs.~\ref{eq:mm0_fsi} and ~\ref{eq:mm_pim_fsi}, the resulting shape of the missing mass distributions is determined by (i) the value of $\varepsilon$ and its distribution among the events in the sample, or in other words, on the probability density function $\xi(\varepsilon)$, (ii) the type of affected final hadron, and (iii) the reaction kinematics. The examples given below are intended to demonstrate the interplay of these factors. These examples test several types of the probability density function $\xi(\varepsilon)$ and consider separately two types of affected hadrons, i.e. proton and positive pion.

\subsection{Illustrating examples}

\vspace{-0.325em}

First, in order to estimate the sensitivity of the missing quantities to hadron momentum alterations, it is convenient to consider the case when the value of $\varepsilon$ is fixed and different from one for all events in the sample. Under this assumption one can examine the distributions of the missing quantities for different fixed values of $\varepsilon$ considering separately the two types of affected final hadrons (the positive pion and the proton).

\afterpage{\clearpage}
\begin{figure}[htp]
\begin{center}
\framebox{\includegraphics[width=\textwidth]{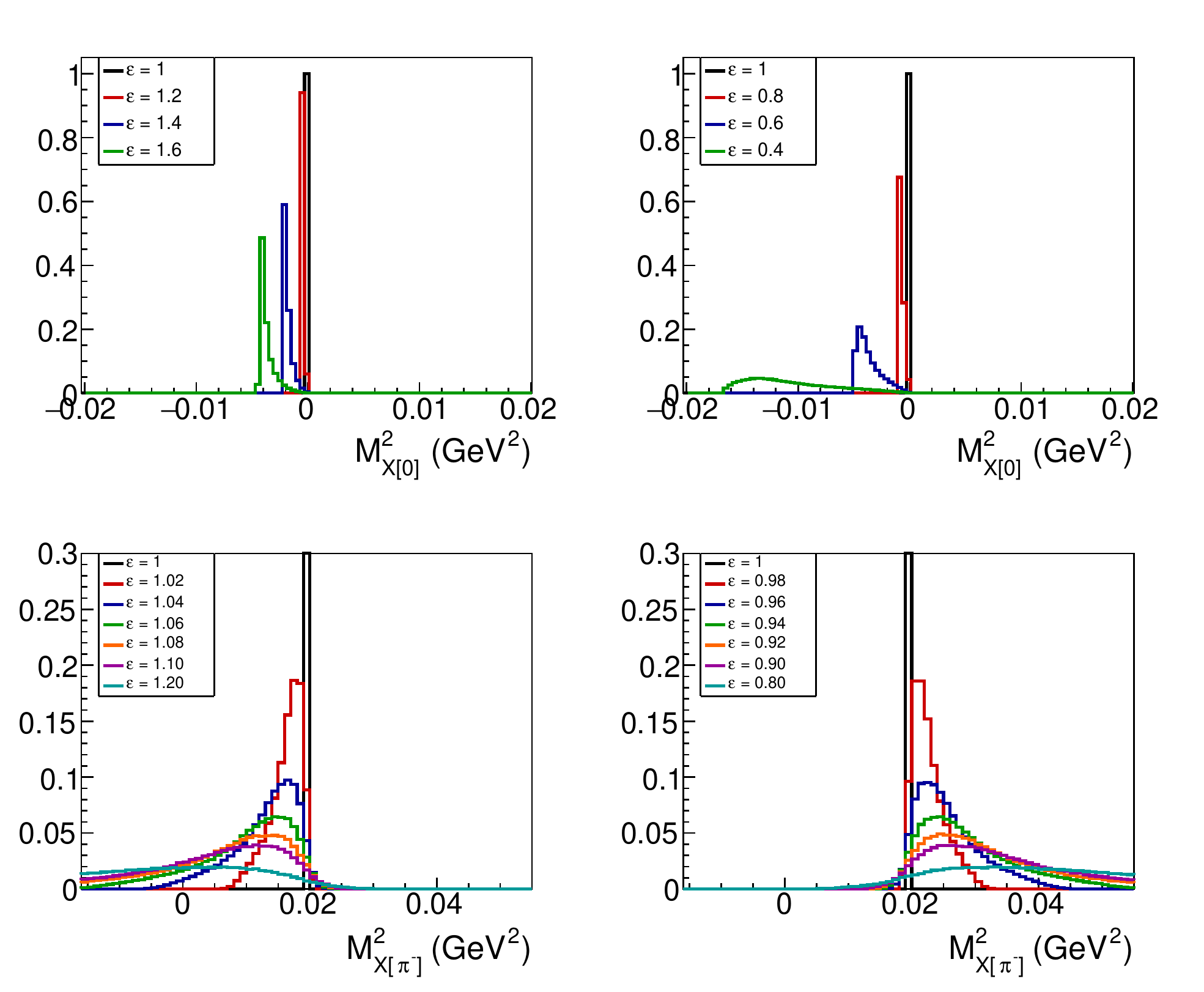}}
\caption{\small Multi-colored histograms show the distributions of $M_{X[0]}^{2}$ (upper panels) and $M_{X[\pi^{-}]}^{2}$ (lower panels) plotted assuming that the $\pi^{+}$ momentum magnitude changes as $p'_{\pi^{+}} = \varepsilon p_{\pi^{+}}$ for all events in the sample. Different colors correspond to different values of $\varepsilon$. Black histograms are given as a reference and correspond to the case when no momentum alterations happen (i.e. $\varepsilon = 1$ for all events). All shown histograms contain equal numbers of events and are normalized in a way that the maximum of the $\varepsilon=1$  histogram is equal to one. The histograms that correspond to the same missing quantity have identical binning. For the quantity $M_{X[\pi^{-}]}^{2}$, shown in the lower panels, note the following, (i) the distributions are zoomed in on small $y$ and (ii) the change in $\varepsilon$ values between the magenta and cyan histograms is five times larger than for all other neighboring histograms. } \label{fig:mm_pip_fsi}
\end{center}
\end{figure}

\begin{figure}[htp]
\begin{center}
\framebox{\includegraphics[width=\textwidth]{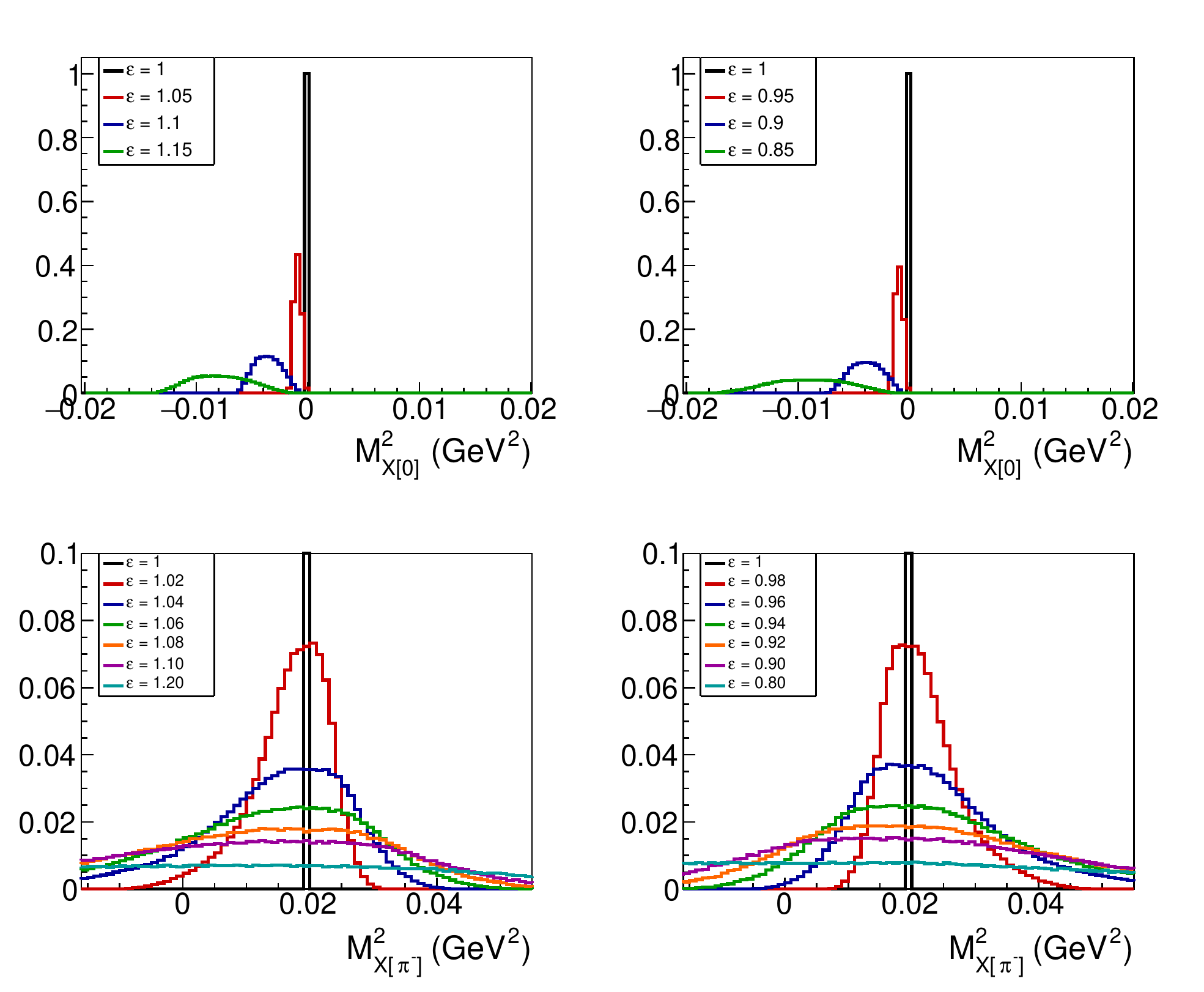}}
\caption{\small Multi-colored histograms show the distributions of $M_{X[0]}^{2}$ (upper panels) and $M_{X[\pi^{-}]}^{2}$ (lower panels) plotted assuming that the proton momentum magnitude changes as $p'_{p'} = \varepsilon p_{p'}$ for all events in the sample. Different colors correspond to different values of $\varepsilon$. Black histograms are given as a reference and correspond to the case when no momentum alterations happen (i.e. $\varepsilon = 1$ for all events). All shown histograms contain equal numbers of events and are normalized in a way that the maximum of the $\varepsilon=1$ histogram is equal to one. The histograms that correspond to the same missing quantity have identical binning. For the quantity $M_{X[\pi^{-}]}^{2}$, shown in the lower panels, note the following, (i) the distributions are zoomed in on small $y$ and (ii) the change in $\varepsilon$ values between the magenta and cyan histograms is five times larger than for all other neighboring histograms. } \label{fig:mm_pr_fsi}
\end{center}
\end{figure}

The multi-colored histograms in Fig.~\ref{fig:mm_pip_fsi} show the distributions of $M_{X[0]}^{2}$ (upper panels) and $M_{X[\pi^{-}]}^{2}$ (lower panels) considering the $\pi^{+}$ to be the affected hadron, which means that all positive pions in the sample change their momenta as $p'_{\pi^{+}} = \varepsilon p_{\pi^{+}}$. Various histogram colors correspond to different values of $\varepsilon$, which are specified in the plots. The quantity $M_{X[0]}^{2}$ is plotted for the sizable deviations of $\varepsilon$ from unity (for $\varepsilon$ from 0.4 to 1.6 in increments of 0.2), since it turns out to be rather insensitive to the change of pion momenta. The quantity $M_{X[\pi^{-}]}^{2}$, being more sensitive to the $\pi^{+}$ momentum change, is plotted for $\varepsilon$ from 0.90 to 1.10 with 0.02 increments and also for $\varepsilon$ of 0.8 and 1.2.

The multi-colored histograms in Fig.~\ref{fig:mm_pr_fsi} show the distributions of $M_{X[0]}^{2}$ (upper panels) and $M_{X[\pi^{-}]}^{2}$ (lower panels) considering the proton to be the affected hadron, which means~that all final protons in the sample change their momenta as $p'_{p'} = \varepsilon p_{p'}$. Various colors~correspond to different values of $\varepsilon$, which are specified in the plots. The quantity $M_{X[0]}^{2}$, being moderately sensitive to the proton momentum change, is plotted for $\varepsilon$ from 0.85 to 1.15 in increments of 0.05, while the quantity $M_{X[\pi^{-}]}^{2}$, being quite sensitive to the proton momentum change, is plotted for $\varepsilon$ from 0.90 to 1.10 with 0.02 increments and also for $\varepsilon$ of~0.8~and~1.2.  

\everypar{\looseness=-1}
The black histograms in Figs.~\ref{fig:mm_pip_fsi} and~\ref{fig:mm_pr_fsi} are given as a reference and correspond to the case when no hadron momentum alterations occur ($\varepsilon = 1$ for all events). All histograms in these figures have equal numbers of events and are normalized in a way that the maximum of the reference histogram is equal to one. The histograms that correspond to the same missing quantity have identical binning in both figures. For the quantity $M_{X[\pi^{-}]}^{2}$, note the following: (i) the distributions are zoomed in on small $y$ to demonstrate their structure and (ii) the change in $\varepsilon$ values between the magenta and cyan histograms is five times larger than for all other neighboring histograms.

\everypar{\looseness=-1}
Examination of the plots in Figs.~\ref{fig:mm_pip_fsi} and \ref{fig:mm_pr_fsi} allows for several interesting conclusions to be made. First, the quantity $M_{X[\pi^{-}]}^{2}$ turns out to be far more sensitive to hadron momentum alterations than $M_{X[0]}^{2}$. Besides this, both missing quantities turn out to be more sensitive to changes in the proton momentum than to changes in the pion momentum. Specifically, for $M_{X[\pi^{-}]}^{2}$, when the same values of $\varepsilon$ are considered for both types of affected hadrons, the distributions plotted for the case of affected protons are about three times lower in height and correspondingly significantly larger in spread than their analogues for the case of affected pions\footnote[8]{Note that as the $\varepsilon\!=\!1$ histogram is always one bin wide, the global normalization of the colored~histograms is bin width specific. The relation of the latter to each other is meanwhile stable, and hence informative.}.

In addition to that, the quantity $M_{X[\pi^{-}]}^{2}$ demonstrates a very peculiar feature: being plotted for the same set of $\varepsilon$ values, the $M_{X[\pi^{-}]}^{2}$ distributions form absolutely different patterns for the two types of affected hadrons. Specifically, when the $\pi^{+}$ is affected, the distributions demonstrate a clear one-sided agglomeration with respect to the position of $m_{\pi^{-}}^{2}$, which is left-sided for $\varepsilon > 1$ and right-sided for $\varepsilon < 1$. Meanwhile, when the proton is the affected hadron, the distributions are widely spread at both sides of $m_{\pi^{-}}^{2}$ for both cases of $\varepsilon > 1$ and $\varepsilon < 1$.

\everypar{\looseness=-1}
This effect can be understood upon closer examination of the final expression in Eq.~\eqref{eq:mm_pim_fsi}. If $\varepsilon > 1$, the quantity in the curly brackets is positive for all events at the right of $m_{\pi^{-}}^{2}$. Meanwhile, if $\varepsilon < 1$, this quantity is negative for almost all events at the left of $m_{\pi^{-}}^{2}$. This general disposition is valid for both types of affected hadrons. However, the interrelation between the momentum and energy terms in the curly brackets turns out to differ for pions and protons. Specifically, for the same values of $\overrightarrow{p}_{\!h}$ for the cases of $h=\pi^{+}$ and $h=p'$, the values of the momentum term are identical, while the absolute value of the energy term for protons is systematically lower than for pions due to the larger mass of the former. As a result, the distributions plotted for the case of affected protons contain more events at the right of $m_{\pi^{-}}^{2}$ for $\varepsilon>1$ (and vice versa, at the left of $m_{\pi^{-}}^{2}$ for $\varepsilon<1$), if compared to the case of pion momentum alterations.

It is also noteworthy that for small values of $p_{h}$, the quantity $(E'_{h}~-~E_{h})$ in Eq.~\eqref{eq:mm_pim_fsi} gives almost indistinguishable results for $h=\pi^{+}$ and $h=p'$. However, the mismatch builds up with the momentum increase. This is due to the fact that for low-momentum hadrons the non-relativistic approximation is applicable, which suppresses the dependence of the considered quantity on the hadron mass. This means that events at the right of $m_{\pi^{-}}^{2}$ for $\varepsilon > 1$ (as well as at the left of $m_{\pi^{-}}^{2}$ for $\varepsilon < 1$) mainly correspond to non-relativistic affected hadrons. Meanwhile, in the considered event sample, protons are mostly non-relativistic, while the majority of pions is relativistic, which reflects the observed difference in the population of the aforementioned areas in the $M_{X[\pi^{-}]}^{2}$ distributions.

Also note that the $M_{X[\pi^{-}]}^{2}$ distributions that correspond to symmetric deviations of $\varepsilon$ from unity, although visually mostly looking as symmetric reflections of each other, have nevertheless asymmetric event allocation. The asymmetry mainly originates from the contribution of the second term in the final expression of Eq.~\eqref{eq:mm_pim_fsi}, which, being always negative, tosses extra events to the left of $m_{\pi^{-}}^{2}$ for all $\varepsilon$ values. As the deviation of $\varepsilon$ from unity grows, the asymmetry grows as well, being accompanied by the gradual shift of the distribution maxima to the left for both options of $\varepsilon > 1$ and $\varepsilon < 1$. The intensity of this effect differs for the two types of affected hadrons: for the case of $h=\pi^{+}$, the effect is truly insignificant for all considered here $\varepsilon$ values, while for $h=p'$ the asymmetry in the event allocation becomes noticeable for deviations of $\varepsilon$ from unity of more than $\sim$~10\%.

With the sensitivity of the missing quantities to hadron momentum alterations now better understood, one can turn to the case when the value of $\varepsilon$ can vary in a certain range among events in the event sample, for example, in the range from 0.8 to 1.2 in increments of 0.01, which results in 41 $\varepsilon$ values in total. The portion of events that carry each value of $\varepsilon$ is then determined by the probability density function $\xi(\varepsilon)$.

Below the shape of the missing quantities is examined, assuming different types of the probability density function $\xi(\varepsilon)$. As the quantity $M_{X[0]}^{2}$ was found to have little sensitivity to changes in the hadron momentum, only the quantity $M_{X[\pi^{-}]}^{2}$ will be further illustrated with the two types of affected hadrons (proton and positive pion) separately considered.

\begin{figure}[htp]
\begin{center}
\framebox{\includegraphics[width=\textwidth]{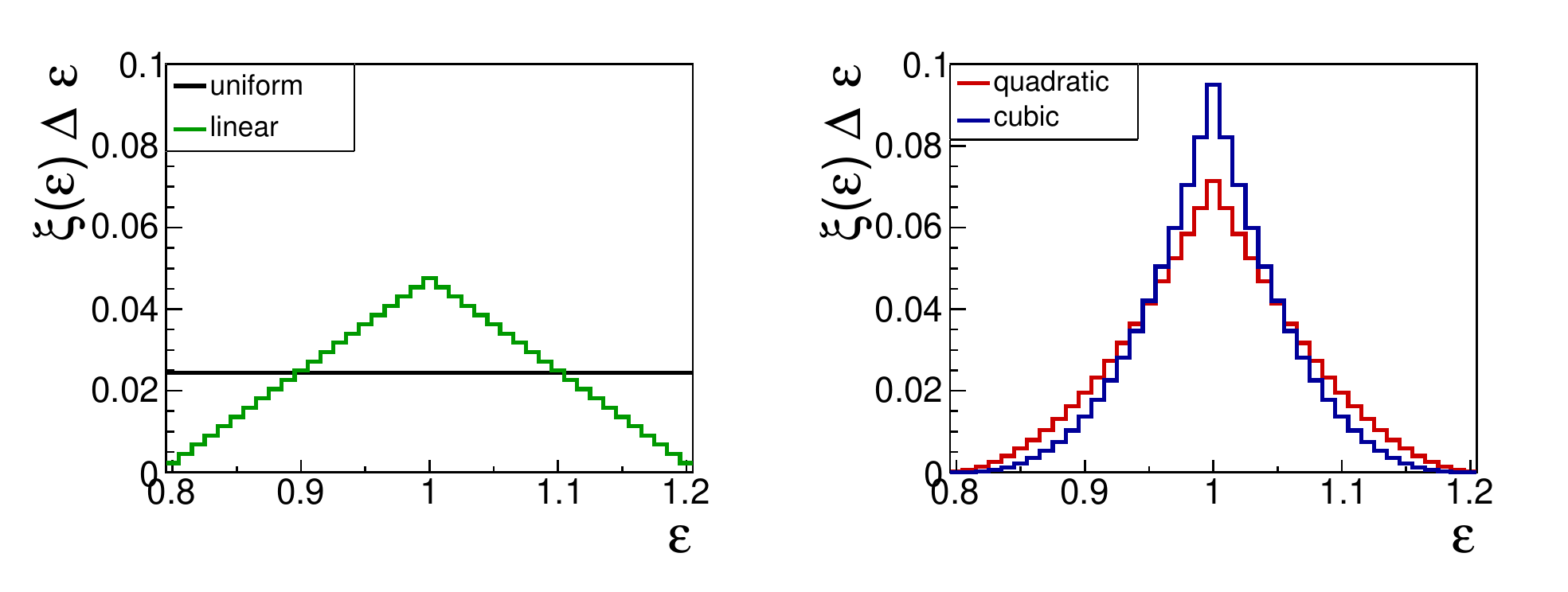}}
\caption{\small Distributions of the quantity $\xi(\varepsilon)\Delta\varepsilon$ for four types of the probability density functions $\xi(\varepsilon)$ considered in this study, i.e. uniform (black), linear (green), quadratic (red), and cubic (blue). The value of $\varepsilon$, which represents the degree of hadron momentum alterations, is assumed to vary in the range from 0.8 to 1.2 in increments of $\Delta\varepsilon=0.01$. Each  $\xi(\varepsilon)$ is normalized to unity. The values on the $y$-axis reflect the portion of events in the sample that carry a particular value of $\varepsilon$.  } \label{fig:density}
\end{center}
\end{figure}

In this study, the following four probability density functions $\xi(\varepsilon)$ are tested: uniform, linear, quadratic, and cubic. The corresponding distributions of the quantity $\xi(\varepsilon)\Delta\varepsilon$ are shown in Fig.~\ref{fig:density}, where $\Delta\varepsilon=0.01$ is the increment of the considered $\varepsilon$ values. Each $\xi(\varepsilon)$ is normalized to unity. The values on the $y$-axis reflect the portion of events in the sample that correspond to a particular value of $\varepsilon$. For simplicity, all tested here probability density functions are taken to be symmetric with respect to $\varepsilon=1$, which assumes equal portions of events with the same percentage of the momentum gain and loss to be present in the sample.

The uniform probability density function corresponds to the black histogram on the left panel of Fig.~\ref{fig:density} and represents the case when equal event portions carry various values of $\varepsilon$. In this example, each of 41 considered $\varepsilon$ values acquires a portion of about 2\% of events in the sample. The uniform probability density function, although hardly being physical, gives the advantage of impartial non-weighted judging of isolated contributions of events with different values (or subranges) of $\varepsilon$ to the resulting distributions of the missing quantities.

\begin{figure}[htp]
\begin{center}
\framebox{\includegraphics[width=\textwidth]{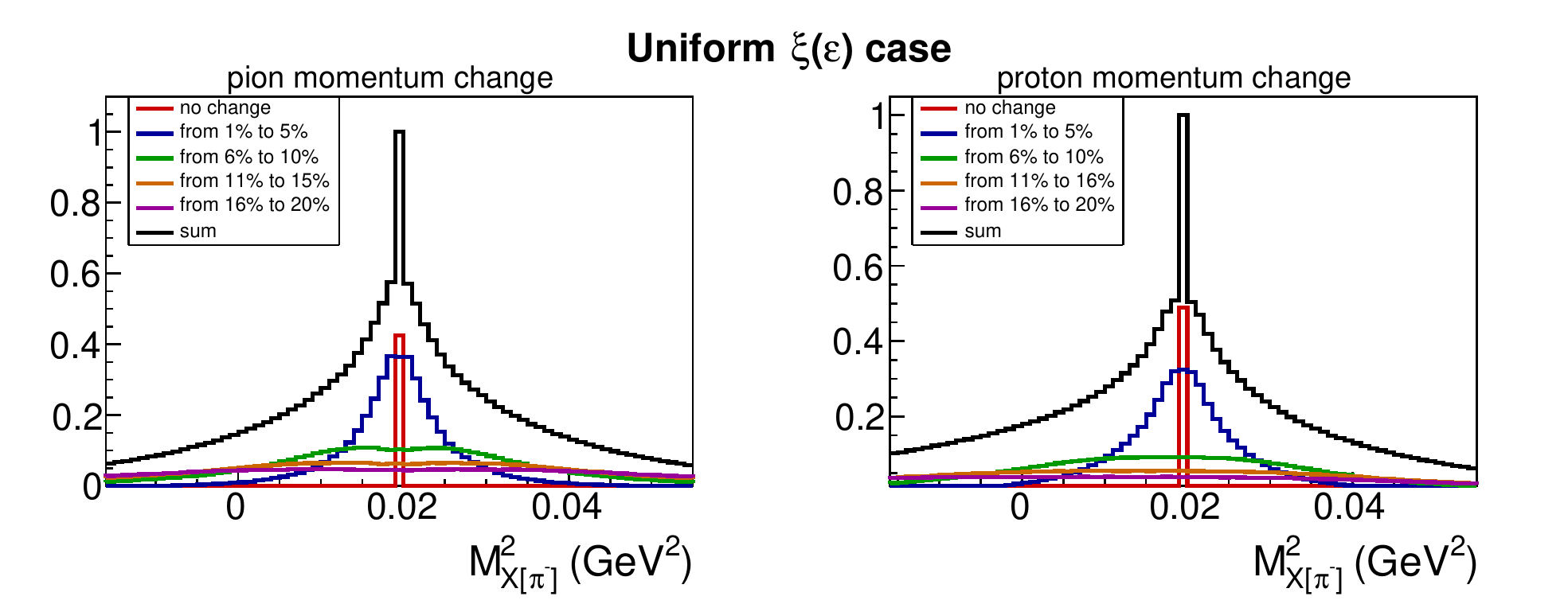}}
\caption{\small Distributions of $M_{X[\pi^{-}]}^{2}$ considering the uniform distribution of $\varepsilon$ values for two types of affected final hadrons, i.e. the positive pion (left) and the proton (right). The multi-colored histograms show the contributions from different subranges of $\varepsilon$ (specified in the plots), while the black histogram shows their sum. Note that the red histogram corresponds to a single value of $\varepsilon = 1$, while each other colored histogram incorporates ten values of $\varepsilon$, which means that the red histogram contains ten times less events than each other colored histogram.} \label{fig:mm_fsi_detail}
\end{center}
\end{figure}
\newpage

Figure~\ref{fig:mm_fsi_detail} shows the distributions of $M_{X[\pi^{-}]}^{2}$ considering a uniform distribution of $\varepsilon$ values for two types of affected hadrons, i.e. the positive pion (left) and the proton (right). The multi-colored histograms show the contributions from different subranges of $\varepsilon$, while the black histogram shows their sum. Note that the red histogram, which corresponds to the absence of momentum alterations, includes just one value of $\varepsilon = 1$, while each other colored histogram incorporates ten values of $\varepsilon$. Assuming the uniform $\xi(\varepsilon)$, this means that the red histogram contains ten times less events than each other colored histogram.

As seen in Fig.~\ref{fig:mm_fsi_detail}, events with small momentum alterations ($\lesssim$ 5\%), although acquiring considerable smearing, still form a distinct peak at the position of $m_{\pi^{-}}^{2}$. However, events with larger momentum alterations ($\gtrsim$ 5\%) lose their peaked structure and tend to form a flat background, which resembles the background observed in experimental distributions of $M_{X[\pi^{-}]}^{2}$ for the reaction occurring off a proton bound in a deuteron~\cite{Skorodumina:2015rea,skorodum_an_note:2019}.

In Fig.~\ref{fig:mm_fsi_detail} one more interesting feature can be observed. Specifically, for the proton momentum alterations (right panel), the left tail of the total distribution (black histogram) contains noticeably more events than its right tail, whereas for the $\pi^{+}$ momentum alterations, both tails look equally populated (left panel). This effect originates from the described above asymmetry in the allocation of events that correspond to symmetric deviations of $\varepsilon$~from~unity, which is more pronounced for the case of affected protons.


\begin{figure}[!ht]
\begin{center}
\includegraphics[width=\textwidth]{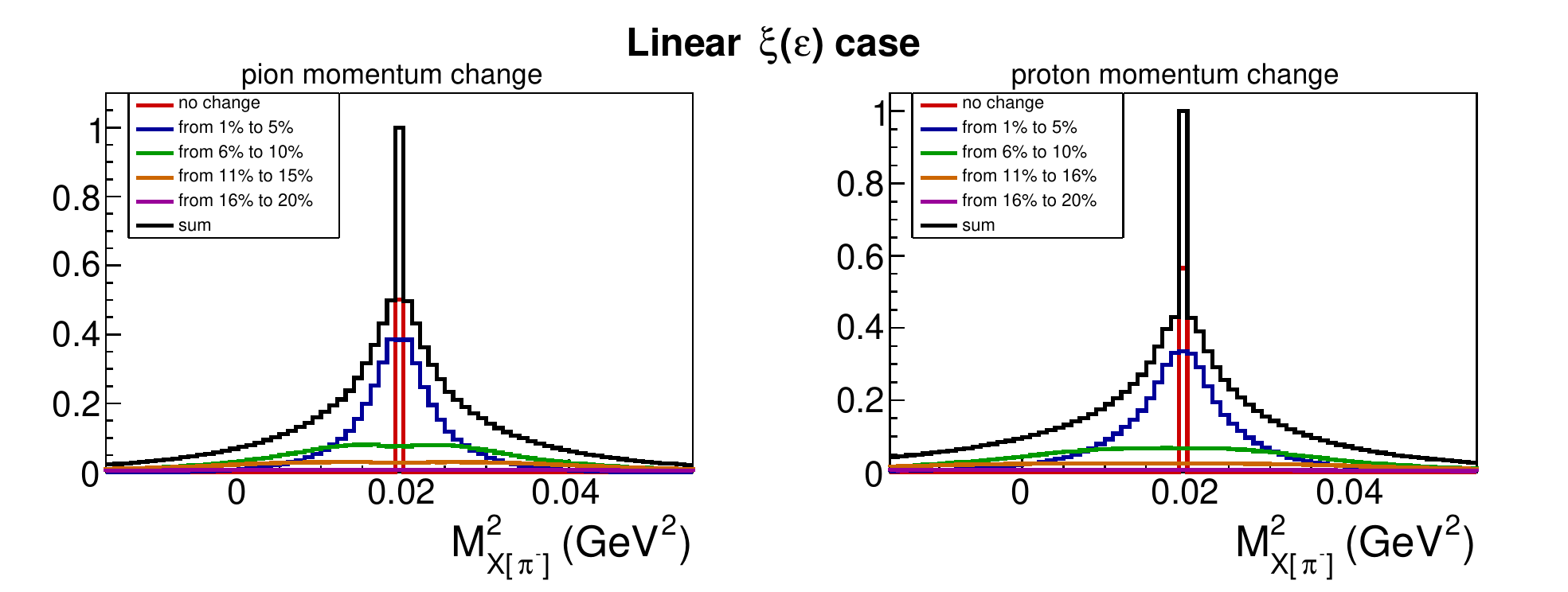}
\includegraphics[width=\textwidth]{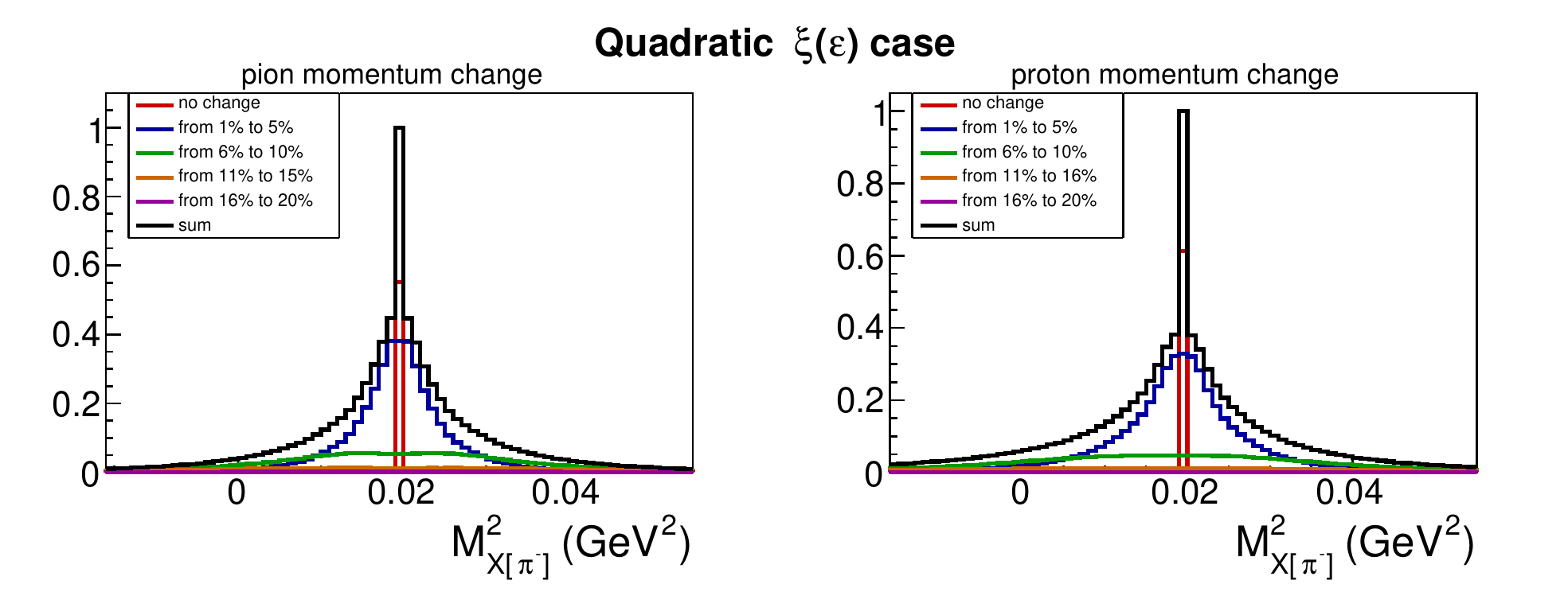}
\includegraphics[width=\textwidth]{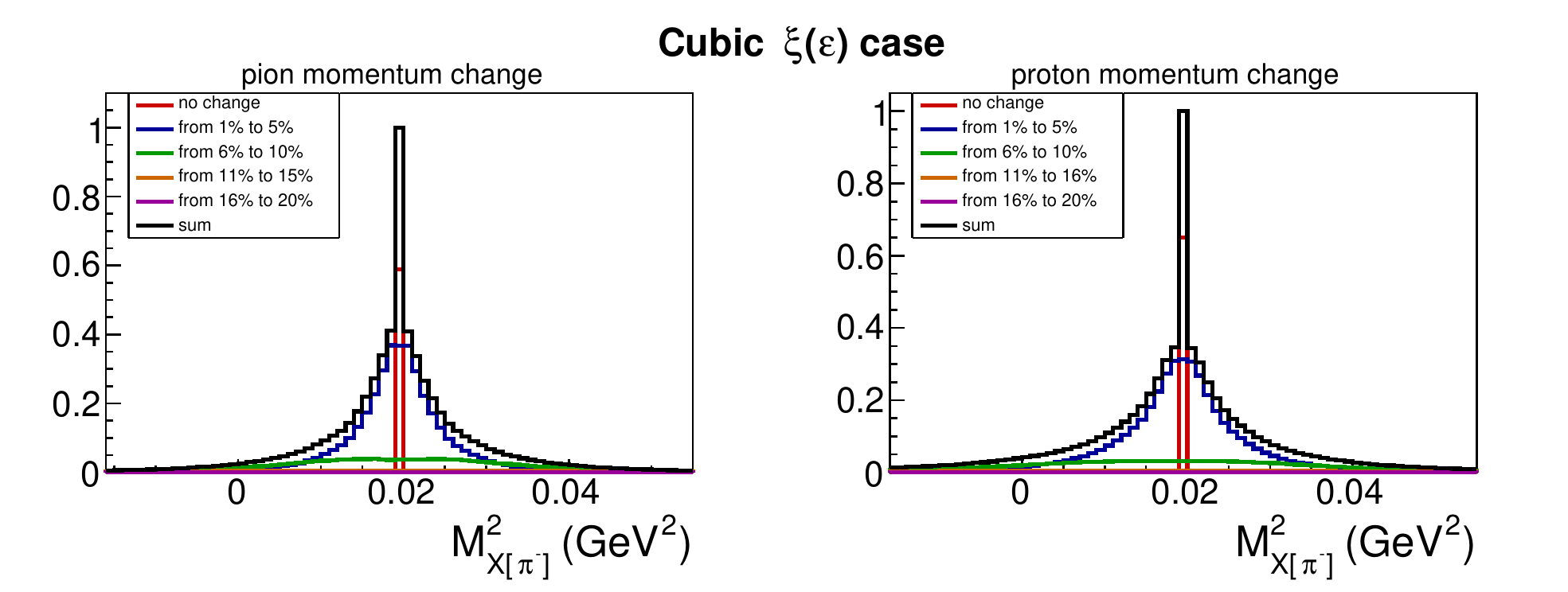}
\end{center}
\vspace{-0.6cm}
\caption{\small Distributions of $M_{X[\pi^{-}]}^{2}$ considering the non-uniform distributions of $\varepsilon$ values for two types of affected final hadrons, i.e. the positive pion (left) and the proton (right). The cases of linear, quadratic, and cubic probability density functions are shown from top to bottom. The multi-colored histograms show the contributions from different subranges of $\varepsilon$ (specified in the plots), while the black histogram shows their sum. }
\label{fig:fsi_combined}
\end{figure}

Now one can employ three non-uniform probability density functions: linear, quadratic, and cubic, which correspond to the green, red, and blue curves in Fig.~\ref{fig:density}, respectively. All of them imply the gradual probability decline with growing momentum deviations, letting events with no momentum alterations to form the largest portion in the sample. Meanwhile, the size of the unaffected event portion and the degree of the probability decline are different for each function.

Figure~\ref{fig:fsi_combined} shows the distributions of $M_{X[\pi^{-}]}^{2}$ plotted employing the aforementioned non-uniform density functions $\xi(\varepsilon)$ for the two types of affected hadrons, i.e. the positive pion (left) and the proton (right). The cases of linear, quadratic, and cubic probability density functions are shown from top to bottom. Again, the multi-colored histograms show the contributions from different subranges of $\varepsilon$, while the black histogram shows their sum.

From Fig.~\ref{fig:fsi_combined} one can conclude that for a probability density function that peaks at $\varepsilon = 1$ and gradually declines with growing degree of momentum alterations, the cumulative contribution from large momentum alterations ($\gtrsim$~10~\%) to the total distribution thins to imperceptible values. The disturbances of the resulting distributions then do not form substantial flat background, staying on a level of smearing.

\afterpage{\clearpage}

%% file: text/concl/concl.tex
\newpage
\chapter{Conclusion}
\mbox{}\vspace{-\baselineskip}

Peculiar features of missing mass distributions were investigated by close examination of the quantities $M^{2}_{X[\pi^{-}]}$ and $M^{2}_{X[0]}$ for the double-pion electroproduction off protons. The five factors, which typically affect the missing mass distributions in experimental studies of exclusive reactions, were examined in detail, i.e. radiative effects, admixtures from other channels, detector resolution, Fermi smearing, and final state interactions with spectator nucleons. The influence of each factor was investigated individually. The examination is supported by the theoretical calculations and the Monte-Carlo simulation.

\everypar{\looseness=-1}
Various combinations of the considered five factors are thought to cover the vast majority of the situations that can be observed in experiments of meson photo- and electroproduction off protons (including both cases of free and bound protons and different beam energies). This examination is therefore intended to facilitate the challenging task of isolating a particular reaction channel during the data analysis process and hence may be helpful for future CLAS12 experiments.

Beyond that, the modeling of kinematic manifestations of final state interactions with spectator nucleons is attempted in Sect.~\ref{sec:fsi}. This modeling may be of special interest for studies of exclusive reactions off bound nucleons as it represents a step towards the reliable simulation of these complicated and so far not fully explored effects.

If needed for comparison, experimental distributions of the considered missing quantities can be found e.g. in Refs.~\cite{Fed_an_note:2017,Fed_paper_2018,Arjun} for the double-pion electroproduction off free protons and in Refs.~\cite{Skorodumina:2015rea,skorodum_an_note:2019} for the reaction occurring off protons bound in deuterium.